\documentclass[a4paper,floatfix]{iopart}

\usepackage[latin1]{inputenc}
\usepackage[english]{babel}
\usepackage{ae}
\usepackage{color}
\usepackage{graphicx}
\usepackage{subfigure}
\usepackage{hyperref}
\hypersetup{
  colorlinks   = true, 
  urlcolor     = blue, 
  linkcolor    = blue, 
  citecolor   = blue 
}

\begin{document}

\title{High-sensitivity plasmonic refractive index sensing using graphene}

\author{Tobias Wenger$^1$, Giovanni Viola$^1$, Jari Kinaret$^2$, Mikael Fogelstr\"om$^1$, and Philippe Tassin$^2$}

\address{$^1$ Department of Microtechnology and Nanoscience (MC2), Chalmers University of Technology, SE-412 96 G\"oteborg, Sweden}
\address{$^2$ Department of Physics, Chalmers University of Technology, SE-412 96 G\"oteborg, Sweden}

\date{\today}

\begin{abstract}
We theoretically demonstrate a high-sensitivity, graphene-plasmon based refractive index sensor working in the mid-infrared at room temperature. The bulk figure of merit of our sensor reaches values above $10$, but the key aspect of our proposed plasmonic sensor is its surface sensitivity which we examine in detail. We have used realistic values regarding doping level and electron relaxation time, which is the limiting factor for the sensor performance. Our results show quantitatively the high performance of graphene-plasmon based refractive index sensors working in the mid-infrared.

\end{abstract}


\maketitle

\ioptwocol

\section{Introduction}
Surface plasmons are collective charge density oscillations on conducting surfaces which have been used for sensing purposes over the last two decades \cite{Wong2014,Nguyen2015}. Recently, graphene has emerged as a new plasmonic material, active in the terahertz to mid-infrared part of the spectrum \cite{Low2014}. Graphene plasmons combine low losses and large field confinements with a unique external tunability \cite{Chen2012,Fei2012}. This makes them attractive for applications such as modulators \cite{Low2014} and photodetectors \cite{Koppens2014}, but also for chemical sensors and biosensors \cite{Grigorenko2012}. In contrast to other plasmonic materials, such as silver and gold, the ability to gate the graphene sample offers the possibility to create plasmonic devices that selectively probe for different molecules, as well as obtaining broadband spectroscopic fingerprints from biomolecules. Surface plasmon resonances help achieve label-free, high throughput detection and screening of biomolecules for drug discovery, genomics, bioengineering, and environmental monitoring \cite{Wong2014,Nguyen2015} and such systems are expected to have a large impact in the future \cite{Maccaferri2015}.

\begin{figure}[h!]
\centering
\includegraphics[width=0.48\textwidth]{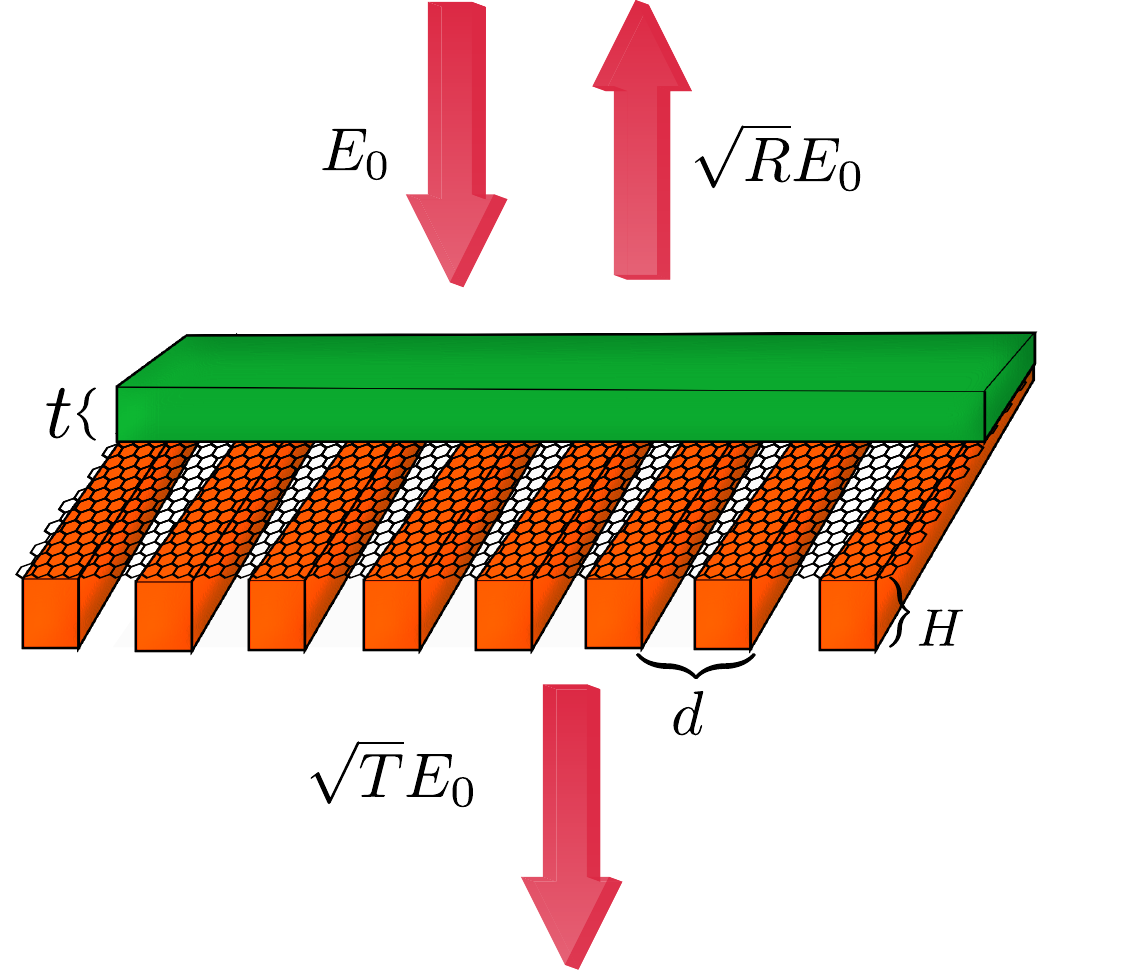}
\caption{\label{fig:setup}Graphene on a subwavelength dielectric grating (orange) with height $H$. On top of the graphene is a substance with refractive index $n$ and thickness $t$. The sensor works by detecting shifts in the plasmon resonance that are induced by the presence of this substance. The substance layer covers the entire graphene surface, in the figure it is made smaller for illustrative purposes. The arrows show the incident, reflected, and transmitted electromagnetic waves. The polarization of all propagating waves is in the direction of the periodicity, i.e. parallel with the distance $d$. Throughout the article we use $d=H=130$ nm and the grating material has a dielectric constant $\varepsilon_r=3$.}
\end{figure}

Previously, graphene nanoribbons \cite{Li2014,Farmer2016,Rodrigo2015} and graphene disks \cite{Marini2015} have been studied for biosensing and chemical sensing purposes. The common feature in these works is the use of graphene plasmons and their field localizing properties to enhance the sensing signal. Refs.~\cite{Li2014,Farmer2016,Rodrigo2015} studied vibrational modes of molecules and it was found that coupling them to graphene plasmons enhances the signal by a factor of $5$ \cite{Li2014} and that a graphene plasmon based sensor can detect minute amounts of gas, down to levels of $50$ zeptomol$/\mu$m$^2$ \cite{Farmer2016}. In addition, Ref.~\cite{Rodrigo2015} demonstrated a gate tunable sensor capable of extracting the permittivity of a molecular layer on the graphene sensor, thereby enabling selective sensing of biomolecules.

In this article we exploit, theoretically, the graphene plasmon frequency shift as a tool to detect small changes in the refractive index surrounding the graphene sheet. To investigate the frequency shift of the plasmons we use the optical reflectance and/or transmittance off graphene on a subwavelength dielectric grating (see Fig.~\ref{fig:setup}). The grating is needed to overcome the large momentum mismatch between the incident light and the graphene plasmons. Previous studies have used a local Random Phase Approximation (RPA) to treat the graphene conductivity. However, a nonlocal approach is known to better approximate the plasmon dispersion as well as incorporate losses in light scattering more correctly \cite{Wenger2016}. In this article we use a nonlocal graphene conductivity obtained from RPA \cite{Wenger2016,Wunsch2006,Hwang2007,Jablan2009}. We also quantify the sensing figure of merit (FoM) of graphene plasmons and examine its dependence on the thickness of the material to be sensed. This is important for sensing small amounts of biomolecules and chemicals on the sensor surface.

\section{Graphene plasmons and damping}
To use graphene plasmons we first determine their frequencies and how the frequency is related to the wavelength. This information is used to make the grating distance match a suitable plasmon frequency for the sensor setup. The graphene plasmon dispersion can be found by solving the non-retarded, $q\gg\omega/c$, dispersion equation \cite{Wenger2016,Jablan2009}
\begin{eqnarray}
\varepsilon_1+\varepsilon_2-\frac{q_1\mathrm{Im} \left [\sigma(q_1,\omega)\right ]}{\omega\varepsilon_0} =0\label{eq:dispReal}\\
\frac{q_2}{q_1}=\frac{\mathrm{Re}[\sigma(q_1,\omega)]}{\frac{\partial}{\partial_{q_1}}(q_1\mathrm{Im}[\sigma(q_1,\omega)])},\label{eq:dispImag}
\end{eqnarray}
where $q=q_1+iq_2$, which also includes the lowest order estimate for the plasmon losses given by $q_2$. In Eq.~\ref{eq:dispReal}, $\varepsilon_1$ and $\varepsilon_2$ are the dielectric constants above and below the graphene sheet and $\varepsilon_0$ is the vacuum permittivity.

\begin{figure}[h!]
\centering
\includegraphics[width=0.48\textwidth]{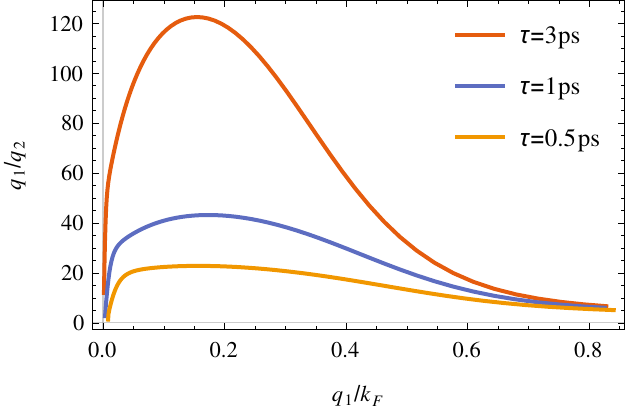}
\caption{\label{fig:damping}Figure showing the calculated inverse damping ratio, $q_1/q_2$, for three different relaxation times. All curves are calculated at room temperature and with a carrier density of $n_e=8\cdot 10^{12}/$cm$^2$.}
\end{figure}

The ratio $q_1/q_2$, sometimes called the inverse damping ratio, is a measure of how many oscillations the plasmon makes before it is damped out completely. For sensing purposes, this number should be as large as possible since this indicates well defined plasmon resonances that can be interacted with. Plasmons in graphene are damped by several sources, such as impurities, phonons, and finite temperature, which affect the performance of the sensor adversely. In this article we include losses through the relaxation-time approximation \cite{Jablan2009,Mermin1970}, in which it is assumed that the combined effect of all electron losses (phonons, impurities etc.) contribute to a total electron relaxation time $\tau$. The total $\tau$ can be estimated from either DC transport measurements or terahertz measurements, and the results in literature for the relaxation time varies between tens of femtoseconds to above one picosecond \cite{Tassin2013}. For concreteness, we set $\tau=1$ ps which can be achieved by boron nitride encapsulation and/or current annealing. To focus the attention we have not explicitly included any encapsulation in this article, but there is no principal problem with encapsulation for the proposed sensor. However, the encapsulation layer must be very thin to allow the substance layer to change the refractive index close to the graphene.

Fig.~\ref{fig:damping} shows the inverse plasmon damping ratio for several different scattering times versus $q_1/k_F$, the inverse grating distance normalized by the Fermi wave vector. All curves are calculated at room temperature and for a carrier density of $n_e=8\cdot10^{12}/$cm$^2$. Large inverse damping ratio indicates that the plasmon is a well defined resonant state. It is clear from the figure that there is an optimum in terms of damping that occurs around $q_1/k_F\approx0.15$. This means that for these specific parameters there is an optimal grating distance to obtain low plasmon damping. The optimum occurs as the temperature damps the plasmons with large $q_1/k_F$ and the electron relaxation damps the plasmons with small $q_1/k_F$, creating the optimum at an intermediate value of $q_1/k_F$.

There is also an intrinsic graphene phonon at an energy of $0.2$ eV, or approximately $50$ THz, for energies above this energy the plasmon becomes severely damped due to the phonon coupling\cite{Low2014,Jablan2009}. This additional damping could in principle be added to the relaxation time but we choose instead to restrict ourselves to energies below the phonon energy. We do this as we anticipate that long relaxation times are essential for the performance of the sensor.

Keeping to energies below $50$ THz, and with the knowledge of the optimal damping, we set $q_1/k_F=0.1$, making the grating distance $d=130$ nm and the plasmon frequency approximately $40$ THz ($\lambda\approx 7.5\ \mu$m). The width of one grating is $d/2$ and the empty space between the gratings also has a width $d/2$. The grating height is set to be $H=130$ nm and the grating material has a dielectric constant $\varepsilon_r=3$. These grating parameters remain fixed throughout the article.

\section{Refractive index sensing}
Fig.~\ref{fig:setup} shows the proposed sensing structure, where graphene is on top of a subwavelength dielectric grating. The substance layer on top of the graphene has a thickness $t$, which is initially set to $t=H=130$ nm. To asses the potential of this graphene structure as a refractive index sensor we calculate its optical scattering properties. This is calculated with a finite-element method solver (COMSOL), in which calculations are performed using one unit cell of the grating with periodic boundary conditions. The top and bottom of the computational domain have absorbing boundary conditions. Graphene enters into the solver as a conducting boundary condition which carries a current induced by the external electric field. The nonlocal real space graphene conductivity $\sigma(x-x',\omega)$ is computed by performing a discrete Fourier transform of the nonlocal momentum space graphene conductivity as $\sigma(x-x',\omega)=\sum_j \sigma(q_j,\omega)e^{iq_j(x-x')}$ and $q_j=2\pi j/d$. The graphene current is obtained by $j(x)=\int  \sigma(x-x',\omega)E(x')dx'$ where the integration is over graphene in the unit cell. Note that it is the grating periodicity $d$ that enables the incoming light to overcome the large momentum mismatch and probe plasmons at large $q$. The incoming light is set to be parallel with the periodicity, $q||E$, in order to probe longitudinal plasmons. The nonlocal momentum space conductivity is calculated using linear response theory \cite{Wenger2016} and a number conserving relaxation-time approximation is used to include the damping \cite{Jablan2009,Mermin1970}.

\begin{figure}
\centering
\subfigure[]{\label{fig:bigRef}
\includegraphics[width=0.48\textwidth]{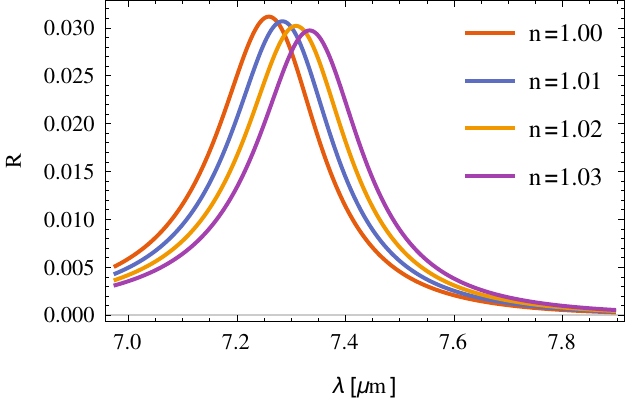}
}
\subfigure[]{\label{fig:bigTrans}
\includegraphics[width=0.48\textwidth]{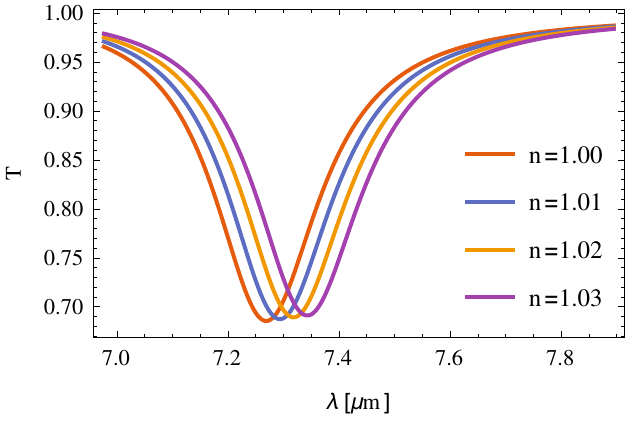}
}
\caption{\label{fig:scattering}Reflectance \textbf{a)} and transmittance \textbf{b)} from the structure for different values of the refractive index on top of the graphene. There is a clear increase in the resonance wavelength as the refractive index increases. The grating dimension is $d=130$ nm, the substance layer thickness $t=130$ nm, and the calculations are done for room temperature.}
\end{figure}

Fig.~\ref{fig:scattering} shows the calculated optical reflectance and transmittance from the graphene structure for different values of the refractive index of the substance on top of the graphene surface. It is clear that the scattering peak in both reflectance and transmittance are sensitive to the refractive index change and the larger the value of the refractive index the larger the increase in wavelength. Note the different scales on the y-axes in Fig.~\ref{fig:scattering}, the height of the reflectance peak is roughly $3\%$ and the dip of the transmittance is approximately ten times larger. The results in Fig.~\ref{fig:scattering} are calculated using a substance layer thickness $t=130$ nm (see Fig.~\ref{fig:setup}).

\begin{figure}[h!]
\centering
\includegraphics[width=0.48\textwidth]{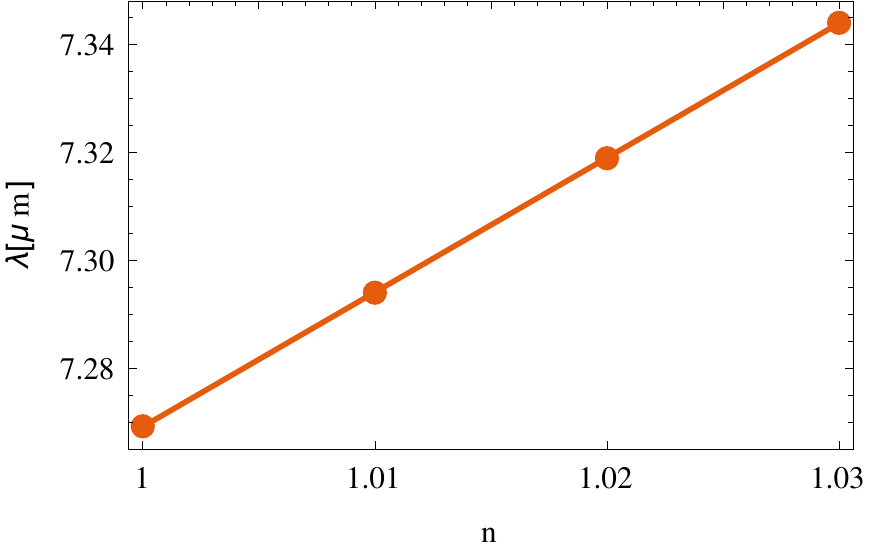}
\caption{\label{fig:sens}Resonance position of the transmittance as a function of refractive index. The dots are the resonance positions extracted directly from the transmittance dips in Fig.~\ref{fig:scattering}. The red line is a linear fit to the data points, obtaining a slope $\partial \lambda/ \partial n=2.5\ \mu$m$/$RIU.}
\end{figure}

To quantify the performance of the proposed refractive index sensor, we compute a sensing FoM. Note that the (free-space) wavelength shift of the resonance is linear in the refractive index change, i.e.,
\begin{equation}\label{eq:bulkShift}
\Delta\lambda = m \Delta n,
\end{equation}
where $m$ denotes the resonance wavelength shift per refractive index unit (RIU) change and has units of meters per RIU. $m$ can be obtained from Eq.~\ref{eq:bulkShift} by $m=\partial\lambda/\partial n$. A commonly used FoM is the so called bulk FoM \cite{Sherry2005,Verellen2011}
\begin{equation}\label{eq:bulkFOM}
{\rm FoM_{bulk} }=\frac{m}{\Gamma_{{\rm FW }}}
\end{equation}
where $\Gamma_{{\rm FW }}$ denotes the full-width at half maximum bandwidth of the resonance. Fig.~\ref{fig:sens} shows the extracted peak positions from the transmittance resonances in Fig.~\ref{fig:scattering} versus the refractive index of the substance layer. By fitting a line through the data points, the slope $\partial\lambda/ \partial n=2.5\ \mu$m$/$RIU is obtained and the resonance widths can be extracted from Fig.~\ref{fig:scattering}. Substituting these values in Eq.~\ref{eq:bulkFOM} we obtain a bulk FoM of $10.9$ for the sensor. 

Comparing with literature, this is a rather large number, meaning that our graphene-plasmon based sensor exhibits a competitive bulk sensing performance. However, bulk FoMs are not always the relevant quantity for the actual merit of sensing applications as pointed out in Refs.~\cite{Li2015a,Li2015b}. The reason is that the bulk FoM is calculated by changing the refractive index of a big volume surrounding the sensor. In biosensing, it is often more important to be sensitive to local refractive index changes close to the surface of the sensor. If the localization of the modes is poor, i.e., if the resonant modes used to probe the index change extend far away from the surface, the bulk FoM can overestimate the performance of the sensor, since it needs a thick layer of refractive index material in order to be as sensitive as the bulk FoM suggests. These considerations make it interesting to quantify the sensor performance for a varying thickness $t$ of the substance layer. To this end, the authors in Refs.~\cite{Li2015a,Li2015b} propose a new way to quantify refractive index sensing using the formula
\begin{equation}\label{eq:shift}
\Delta \lambda = m \Delta n (1 -e^{-2t/L_d})
\end{equation}
where $t$ is the layer thickness of the substance layer, $L_d$ is the decay length, and $m$ is the sensitivity. This formula explicitly includes the layer thickness dependence of the spectral shifts. We point out that this is nothing but a redefinition of the sensitivity $m$, the spectral shifts of the resonances, $\Delta \lambda$, are of course the same. Note that in the limit of layer thickness $t\to\infty$, we regain the bulk expression for the sensitivity $m$, i.e., Eq.~\ref{eq:bulkShift}. Differentiating Eq.~\ref{eq:shift} with respect to $n$ we obtain an equation for the spectral shift per RIU:

\begin{equation}\label{eq:specThick}
\frac{\partial\lambda}{\partial n}= m \left (1 - e^{-2t/L_d}  \right ),
\end{equation}
meaning that the spectral shift per RIU is decreasing with decreasing substance layer thickness $t$.

\begin{figure}[h!]
\centering
\includegraphics[width=0.48\textwidth]{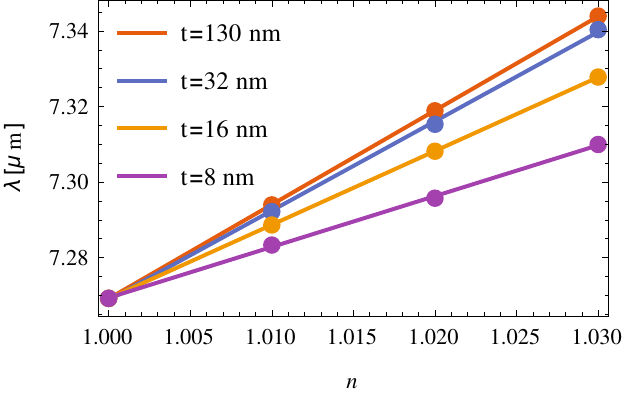}
\caption{\label{fig:sensMany}Figure showing the resonance position as a function of refractive index change, using four different thicknesses $t$ of the substance layer. The red upper line is the same line as in Fig.~\ref{fig:sens}.}
\end{figure}

Fig.~\ref{fig:sensMany} shows the transmittance resonance position as a function of refractive index for different substance layer thicknesses $t$. As expected from Eq.~\ref{eq:shift}, the shift gets smaller as the substance layer thickness is reduced, i.e., the slope of the fitting curves is reduced.

\begin{figure}[h!]
\centering
\includegraphics[width=0.48\textwidth]{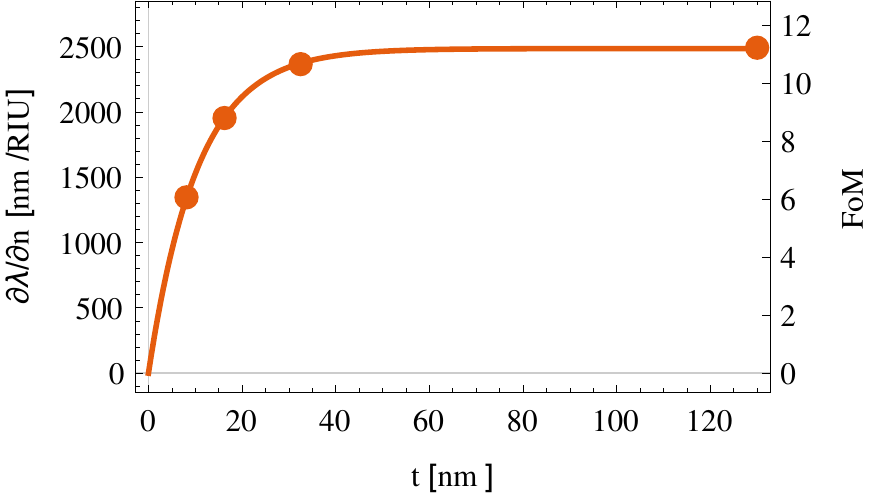}
\caption{\label{fig:thickness}Figure showing the calculated spectral shifts (dots) and a fit to Eq.~\ref{eq:specThick} (solid line). It is clear from this figure that by reducing the substance layer thickness $t$, the resonance shift becomes smaller. For a thickness of $2L_d$ ($42$ nm), the resonance shift per RIU saturates at the bulk value.}
\end{figure}

Fig.~\ref{fig:thickness} (dots) shows the spectral shifts per RIU obtained from the lines in Fig.~\ref{fig:sensMany}. Fig.~\ref{fig:thickness} also shows a fit of this data to Eq.~\ref{eq:specThick} (solid line), obtaining $m=2480$ nm$/$RIU and $L_d=21$ nm. It is clear that the spectral shifts are reduced as the thickness $t$ is reduced, the fit also allows us to extract information about the shifts for very small $t$. Note the axis on the right hand side of Fig.~\ref{fig:thickness}, where the spectral shift is divided by the resonance width, thereby obtaining a thickness dependent FoM. For thicknesses above $2L_d$ ($42$ nm), the FoM reaches the bulk value and below $L_d$ ($21$ nm) the FoM drops to zero. Note that even for thicknesses of a few nanometers the FoM is $2-3$, meaning that the sensor is capable of sensing very thin substance layers on the sensor surface.

The authors in Refs.~\cite{Li2015a,Li2015b} propose to use a second order derivative of the spectral shift described by Eq.~\ref{eq:shift}, giving 
\begin{equation}\label{eq:surfacePrefactor}
\frac{\partial^2 \lambda }{\partial t \partial n}=\frac{2m}{L_d}e^{-2t/L_d}
\end{equation}
which allows the limit $t\to0$ to be taken to extract information about the surface sensitivity defined by the prefactor $\frac{2m}{L_d}$. From the fit in Fig.~\ref{fig:thickness}, $m$ and $L_d$ are extracted to obtain a surface sensitivity prefactor of $2m/L_d=237$. This is a dimensionless number that can be used to compare different sensors. We note that in Refs.~\cite{Li2015a,Li2015b}, surface sensitivity prefactors below $10$ are reported. This highlights the ability of graphene plasmons to strongly localize electromagnetic fields, creating a high sensitivity at the surface.

\section{Discussion}
We have demonstrated that graphene allows for refractive index sensors with better performance than traditional refractive index sensors. We wish to point out that graphene losses limit the sensing FoM. By increasing the graphene quality, it is possible to reach even higher FoMs than reported here. 

Furthermore, temperature plays an important role in limiting the FoM. We have used room temperature throughout this article. However, reducing the temperature increases the FoM and may be interesting for experimental efforts. This increased FoM has two sources, one is that the direct temperature broadening of the single particle continuum is decreased, leading to lower plasmon damping for smaller temperatures. In addition, decreasing the temperature tends to increase the mobility of graphene, this can be understood by considering that lowering the temperature limits the phonon scattering. This leads to a longer electron relaxation time $\tau$, which as previously stated is a main source of plasmon damping in our model. We find that by increasing to a more optimistic estimate for the relaxation time of $\tau=3$ ps (throughout the article we have used $\tau=1$ ps) we obtain a bulk FoM above $20$. In the limit of pristine graphene (i.e., an infinite scattering time, but still at room temperature and nonlocal effects included), we obtain a bulk FoM of roughly $40$, which provides a best-case scenario for the FoM, showing the potential of ultra-sensitive graphene-plasmon based refractive index sensors. Note that the temperature and the relaxation time both enter the equations as fractions of the Fermi energy. Thus, by increasing the Fermi energy and keeping the absolute temperature and relaxation time fixed, their effects on the sensitivity become smaller.

By incorporating a metal gate, the Fermi energy can be tuned and thus changing the resonance frequency of the plasmons. By making small changes to the gate voltage (small enough to avoid substantially increasing the effective size of the temperature and relaxation time) the sensor can sweep over frequency in order to take a spectroscopic fingerprint in the mid-infrared. This allows for selective sensing of molecules in this frequency range.

Even though the bulk FoM of our graphene-plasmon based sensor is large, it is not the largest achieved in the literature \cite{Sadeghi2016}. However, we emphasize that the main strength of our plasmon sensor is its ability to sense refractive index changes that occur very close to the surface. This is due to the strong electricomagnetic field localization facilitated by the plasmons and it is the reason for the enhanced sensing properties of graphene based devices in, for instance, Refs.~\cite{Li2014,Farmer2016,Rodrigo2015,Marini2015}. In this article, we have taken a more general view and calculated the quantitative performance of graphene-plasmon based refractive-index sensing devices. 

Using phase sensitive measurements it is possible to enhance the FoM \cite{Maccaferri2015,Lodewijks2012}. However, this requires a more complicated setup to perform phase measurements of the reflected and/or transmitted light. We note that the phase of the reflected light in our calculations has a pronounced step as the frequency sweeps across the plasmon resonance and the shape of this step is quite robust against losses. Such steps in phase are the basis for the FoM enhancements in Refs.~\cite{Maccaferri2015,Lodewijks2012}. This makes the prospect of phase sensitive measurements together with graphene plasmons very interesting for future investigations.

As a final comment, we emphasize that the square subwavelength dielectric grating was chosen for its simplicity. It is very possible, even likely, that further optimization of this geometry will lead to even better performances of graphene-plasmon based sensors. An interesting possibility is to combine the dielectric structures that we have investigated with metallic nanoparticles used in traditional refractive index sensors.

\section{Conclusions}
We have theoretically demonstrated a graphene-plasmon based refractive index sensor working in the mid-infrared, exhibiting a large bulk FoM of 11. We have thoroughly investigated the performance of the sensor by performing calculations for a reduced thickness of the substance layer. We show that even for a substance layer as thin as a few nanometers the sensor has a FoM of $2-3$, making it possible to sense very small refractive index changes close to the sensor surface. This sensitivity comes from the ability of the graphene plasmons to localize the electromagnetic fields very close to the graphene surface. For distances above roughly $40$ nm the sensor achieves its bulk FoM, meaning that the electromagnetic fields are well localized within this region. The performance of the sensor is limited by electron relaxation and finite temperature. Reducing either the temperature or the scattering, the performance of the sensor may be further improved. We believe that our quantitative findings support the high potential of graphene-plasmon based refractive index sensors in the mid-infrared. 

\section*{Acknowledgements}
The authors thank the Knut and Alice Wallenberg Foundation (KAW) for financial support.

\section*{References}

\bibliographystyle{unsrt}
\bibliography{refs}

\end{document}